\documentclass[aps,pra,twocolumn,floatfix]{revtex4} 
\usepackage[dvips]{graphicx}
\usepackage{float}
\usepackage{bm}
\usepackage{color}
\begin{document}
\title{Equilibrium trapping of cold atoms using dipole and radiative forces
in an optical trap}
\author{Taro Mashimo}
\author{Masashi Abe}
\author{Satoshi Tojo}
\email{tojo@phys.chuo-u.ac.jp}
\affiliation{Department of Physics,
Chuo University, Kasuga, Bunkyo-ku, Tokyo 112-8551, Japan} 
\date{\today}

\begin{abstract}
We report on highly effective trapping of cold atoms
by a new method for a stable single optical trap
in the near-optical resonant regime.
An optical trap with the near-optical resonance condition
consists of not only the dipole but also the radiative forces,
while a trap using a far-off resonance dominates only the dipole force.
We estimate a near-optical resonant trap 
for ultracold rubidium atoms
in the range between $-0.373$ and $-2.23$ THz from the resonance.
The time dependence of the trapped atoms indicates
some difference of the stable center-of-mass positions
in the near-optical resonant trap, 
and also indicates that the differences are caused by 
the change of the equilibrium condition of the optical dipole and radiative forces.
A stable position depends only on laser detuning 
due to the change in the radiative force; 
however, the position is ineffective against the change in the laser intensity,
which results in a change in the radiative force.

\end{abstract} 

\maketitle

\section{Introduction}

Optical control of particles and materials is 
one of the powerful techniques for investigating 
novel phenomena of media for precise observations in highly
qualified isolations \cite{Grier}.
Because of the demonstration of the experimental observation using the
optical control for atoms \cite{ChuPRL},
optical tweezer techniques are applied to broad research fields  \cite{Miller,Gustavson},
dielectric and metal particles \cite{AshkinOL,Svobada},
and viruses and bacteria \cite{AshkinSci}.
Moreover, these techniques as optomechanical engineering have been used for 
realizing precise allocation of atoms \cite{AshkinPNAS},
levitation of nanoparticles to investigate isolated thermal phenomena
\cite{MahamdehOL}, etc.
Nevertheless,
these optomechanics have been applied using the dipole forces only.
The radiative force 
is not preferred because its scattering effect induces
dephasing of quantum states, 
degradation of observation precision,
and heating due to spontaneous emission with absorption
\cite{Metcalf}.

For fundamental research studies, 
optical forces have been developed in the research field of cold atoms
because
sufficient low density allows collision of atoms in the two-body problem
so that experiments are in excellent agreement with theoretical predictions.
A far-off resonant trap (FORT) has been investigated and applied for
research fields involving cold-atom experiments \cite{GrimmAMOP}.
The FORT, which has large detuning from the resonance with more than
tens of THz typically, can hold cold atoms and quantum degenerates in any spin
state for investigating collision properties \cite{TojoPRA} and 
spinor Bose-Einstein condensates (BECs) without heating and dephasing
the quantum states \cite{StamperKurnRMP}, 
and to change the shapes of the trap potential 
and spatial controls of the atoms \cite{PRLGaunt,NJPLeonard}.
The applications of the researches in the FORT mainly involve the use of
the dipole forces, called as optical {\it dipole force} trap.
Moreover, although the near-optical resonant trap (NORT), which
has a small detuning round of several tens of GHz to several THz, 
has been treated with experimental difficulties
because of heating, scattering, and dephasing 
beyond their perturbative conditions by the spontaneous emission
\cite{Szczepkowicz,Bienaime},
there is a possibility of using not only the dipole force but also the
radiative force as the hybrid force trap.

In this work,
we have investigated the hybrid potential, which consists of
the dipole and radiative forces used in the NORT
regime while detuning in the range of sub THz to several THz from the resonance. 
We evaluate the equilibrium trap conditions regardless of the
radiative effect, and find that 
our experimental results of the equilibrium positions are 
in good agreement with the calculations of the threshold detuning under
the condition of stable hybrid forces in a single optical trap.

\section{Theory}
\subsection{Dipole and radiative forces in the NORT}

The trap beam potential using laser light can be written by the Gaussian
beam optics \cite{Mandel}. The electric field amplitude of the propagating beam
along the $z$-axis near the Rayleigh region is approximated as 
\begin{equation}
E(r,z) =  \frac{E_0}{\sqrt{1+(z/z_0)^2}}\exp \left[ -\frac{r^2}{w^2(z)} \right],
\label{eq1} 
\end{equation}
where $w(z) = w_0 \sqrt{1+(z/z_0)^2}$ is the radius of the beam, 
$w_0$ is the beam waist,
$z_0 = \frac{1}{2}k{w_0}^2$ is the confocal parameter,
and $E_0$ is the amplitude of the electric field at the center of 
the beam waist of the trap laser beam. 
The interaction between the atom and the trap laser beam produces the dipole and 
radiative forces in the atomic cloud in the trap region \cite{Lett}.
\begin{figure}[htbp]
\begin{center}
\includegraphics[width=7cm]{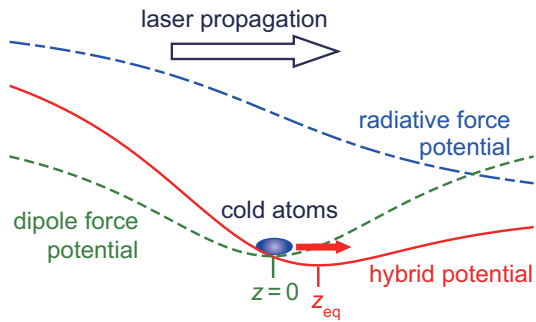}
\end{center}
\caption{(color online).
Geometry of the cold atoms in the NORT.
The trapped atoms located in the focal point
have imbalanced effects by the dipole and the radiative forces
generating the hybrid potential
along with the beam axis of the trap laser.}
\label{potential}
\end{figure}
If 
the detuning of the laser from the atomic resonance $\delta$
is sufficiently larger than
the natural width of the resonance $\Gamma$
and the saturation parameter is as small as
$s=(\mu^2E^2/2\hbar^2)/(\delta^2+\Gamma^2/4) \ll 1$,
the optical dipole and radiative forces at the beam axis
$F_{\rm D}$ and $F_{\rm R}$ are approximated as
\begin{eqnarray}
F_{\rm D} &=&
-\frac{\mu^2}{4\hbar}\frac{{E_0}^2}{\delta}
\frac{2z {z_0}^2}{(z^2+{z_0}^2)^2}
\label{eq2} \\
F_{\rm R} &=&
\frac{\mu^2}{4\hbar}\frac{k_z \Gamma}{2\pi}
\frac{{E_0}^2}{\delta^2}\frac{{z_0}^2}{z^2+{z_0}^2}, 
\label{eq3} 
\end{eqnarray}
where 
$\mu$ is the electric dipole moment of an atom and
$k_z$ is the $z$-component of the wave vector.
In the case of the FORT,
the radiative force is negligible due to $|\delta| \gg k_z z_0 \Gamma/(4\pi)$
\cite{Szczepkowicz}.
Therefore, the trap is called an optical {\it dipole} trap.
However, in the case of the NORT,
there is a possibility of the existence of an equilibrium condition with
variables $\delta$ and $z$ due to the presence of the radiative force. 
As a result,
the dipole and radiative forces generate a hybrid trap potential
and change the position of the potential minimum.
The change in the position due to the hybrid trap forces from
that in the optical dipole force is defined as an equilibrium position.
The equilibrium position, or the displacement from the initial position,
$z_{\rm eq}$ is caused by the balance condition in Eqs.(\ref{eq2}) and (\ref{eq3}),
and is expressed as the solution of the equation
$z^2- 4\pi z|\delta|/(k_z \Gamma)+{z_0}^2=0$ 
with $-z_0 < z < z_0$ as 
\begin{equation}
z_{\rm eq} = \frac{2\pi|\delta|}{k_z \Gamma}-
\sqrt{\left( \frac{2\pi|\delta|}{k_z \Gamma} \right)^2 -{z_0}^2}.
\label{eq4}
\end{equation}
Equation (\ref{eq4}) indicates that 
$|\delta| > k_z^2{w_{0}}^2 \Gamma/(8\pi^2)$ 
in the presence of equilibrium.
Figure \ref{potential} illustrates a typical geometry of the trapped
cold atoms in the NORT.
Owing to the radiative force potential,
the total potential of the trap is modified as the hybrid potential from
the dipole force potential.
The change of the equilibrium position, indicated as $z_{\rm eq}$,
generates variable oscillating motions of the
trapped atoms from the initial position $z=0$.
We note that the equilibrium position is independent of the electric
field amplitude $E_0$.

\subsection{Loading efficiency at a finite temperature in the NORT}

Initial loading rate of a typical magneto-optical trap (MOT)
to an optical trap depends on
the profile of the trap beam, including the effect of gravity and the
temperature of the cold atomic cloud \cite{Dumke}. 
While loading using an optical beam, 
the scholar potential generated by the dipole force determines the
loading efficiency, and the vector potential generated by the radiative
force affects along the beam axis. 
Owing to the depth of the dipole force potential, 
the loaded atoms can be trapped on the axis
in spite of the existence of the vector potential. 
The optical dipole force potential in the case of 
sufficiently small saturation parameter $s \ll 1$ can be expressed as 
\begin{equation}
U_{\rm D} \simeq \frac{\hbar \Gamma^2}{8\delta} \frac{I}{I_{\rm s}}
\label{eq5}
\end{equation}
where $I$ is the laser intensity, and $I_s = 2\pi hc \gamma/3\lambda^3$ 
is the saturation intensity. 
The trap generated by a typical laser beam has a
two-dimensional Gaussian profile on the cross section perpendicular
to the beam axis. 
In the electric field of the trap, 
the trap potential near the Rayleigh region in Eq.~(\ref{eq5})
can be rewritten as  
\begin{equation}
U_{\rm D}(r,z) = \frac{\epsilon_0\hbar \Gamma^2}{16 I_{\rm s} \delta}
\frac{{w_0}^2 {E_0}^2}{w^2(z)} \exp \left[- \frac{2r^2}{w^2(z)} \right]
\label{eq6}
\end{equation}
The net trap potential should be modified with an effect of the gravity as  
$U_{\rm D}' = U_{\rm D}(r,z) - u_{\rm g}(z)$
where $u_{\rm g}(z)$ is the compensated potential due to the gravity. 
The modified potential is determined by the profile of the trap beam.
In the condition of 
a typical confinement of the horizontal trap beam for ultracold atoms,
$u_{\rm g}(z)$ can be approximated to be constant 
as $u_{\rm g0}$~\cite{Shibata}. 

A thermal atomic cloud is generally loaded into the optical trap up to an
energy of 1/10 of the potential depth \cite{Dumke,Shibata}. 
We regard the energy as the threshold limit 
below which the optical trap can
hold lower-velocity atoms in accordance with the Maxwell-Boltzmann
distribution at a finite temperature. 
The trapped number of atoms in unit length at $z$-axis is expressed as 
the cumulative distribution function of the Maxwell-Boltzmann
distribution $h_{\rm MB}(T,U(r))$ from the atoms with the lowest velocity
in the potential $U(r)$ at a temperature of $T$. 
By using modified trap potential with an effect of the gravity, 
we can derive the total number of trapped atoms expressed as 
\begin{equation}
N = n_0 \int_{-z_a}^{z_a} \int_0^{r_a}
h_{\rm MB}(T,U'_{\rm D}(r))2\pi rdrdz,
\label{eq7}
\end{equation}
where $n_0$ is the averaged density of the atomic cloud in the trap 
at a temperature of $T$,
and $r_a$ is the radius of the atomic cloud.

\section{Experimental setup}
\subsection{Cooling setup}

The experimental apparatus and procedure used for ultracold $^{87}$Rb
atoms are almost the same as those described in Ref.~\cite{Shibata},
except for the optical trap system.
We prepare cold $^{87}$Rb atoms in a vacuum glass cell using
a conventional MOT.
After pre-cooling the Rb atoms,
we turn off the magnetic-field gradient after the compression and start
the polarization gradient cooling (PGC), 
which consists of two parts. 
In the first part, the detuning of the cooling laser is swept from $-32$
to $-80$ MHz with a decrease of the initial intensity by 1/4. 
The second part starts with a frequency jump of the laser
locking point from the $F' = 3$ peak to the $F' = 2, 3$ crossover peak. 
This jump shift of $-133$ MHz enables large detuning. 
We use a commercially available laser servo
controller (Vescent Photonics Inc., D2-125),
which enables the frequency jump with a sample-and-hold circuit. 
The repump beam detuning is set
to $-123$ MHz to obtain gray-molasses cooling \cite{Hemmierich,Boiron}.
After the second part,
$1.8 \times 10^8$ atoms at less than 5 $\mu$K are typically produced with
the cancellation of the residual magnetic field using 
three Helmholtz coil pairs for highly effective PGC. 
The coil currents are finely adjusted to minimize the temperature in
the molasses after the cooling in the second part. 
We estimate the residual field to be less than 50 mG.
The optical trap beam is switched on after the atom cloud 
of diameter approximately 400 $\mu$m
is cooled with the cooling and repump beams. 
The molasses and the optical trap beam are overlapped for 65 ms.
In the absence of the overlap, the atoms in the molasses fall off
and are separated from the optical trap.
The loading of atoms is completed in this period.
After loading in the optical trap, the cooling beams and repumping beams are turned off.
We change the holding time from 10 to 300 ms
to increase the hold time by 10 ms. 
At the end of holding time, 
the atoms are irradiated by the weak repumping beam,
and almost all the atoms are pumped to the $F=2$ state. 
The number of atom and the temperature of the atoms in the optical trap
is measured through the absorption imaging after the time of flight.

\subsection{The NORT apparatus}

\begin{figure}[tb]
\begin{center}
\includegraphics[width=8cm]{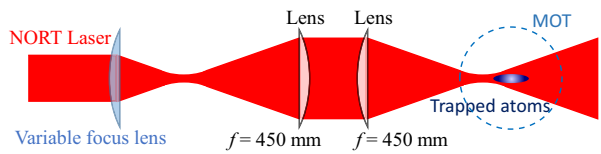}
\end{center}
\caption{(color online).
Experimental geometry of the NORT.
By changing in the focal length of the variable focus lens,
the beam parameters of the NORT can be controlled.
}
\label{setup}
\end{figure}
\begin{figure}[htbp]
\begin{center}
\includegraphics[width=6.5cm]{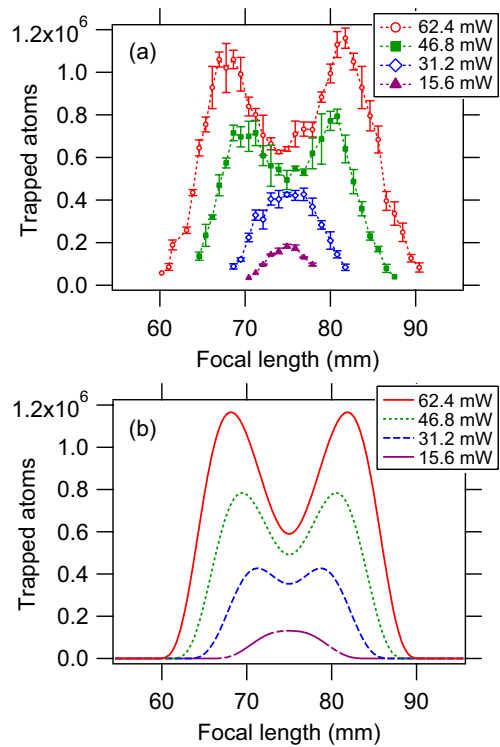}
\end{center}
\caption{(color online).
Focal length dependence of the number of atoms:
(a) Experimental results at
62.4 mW (open circles),
46.8 mW (closed squares),
31.2 mW (open diamonds),
and 15.6 mW (closed triangles).
(b) Calculations with a single fitting parameter to compensate the number
 of atoms at
62.4 mW (solid line),
46.8 mW (dotted line),
31.2 mW (dashed line),
and 15.6 mW (dotted-dashed line).
}
\label{numexpcal}
\end{figure}
Figure \ref{setup} shows the experimental setup of our NORT.
The trap beam was generated from a semiconductor laser, and the detuning was adjusted from the
D$_2$ line from  $-0.37$ to $-2.23$ THz. 
The collimated trap laser beam passed through the expanding lens system, which consisted
of a variable focus lens and two fixed focus lenses with $f = 450$ mm 
after passing through a single-mode optical fiber to clean up 
its spatial mode and to improve the spatial stability of the laser beam. 
The focal length of the variable lens was set at approximately 75 mm to achieve the focus condition at the center of the molasses cloud.
The beam profile can be approximated to be a circle, 
and the focal point
can be displaced by changing the focal length of the variable focus lens
in the range of $f =$ 60 - 90 mm. 
The beam power was set between 7.8 and 62.4 mW 
with linear polarization along the horizontal axis. 
The range of the potential depths was estimated to be from $k_{\rm B}
\times 20$ to $k_{\rm B} \times 300$ $\mu$K.
The volume and depth of the trap were dominant for the loading
efficiency from the ultracold atomic clouds in the optical molasses
condition \cite{Shibata}.
By using a tilted 785-nm band-pass filter, 
we reduce the scattering and heating effects from the resonance of the D$_2$ line
in the NORT.

\section{Results and discussion}

We measure the remaining number of atoms in the NORT for 20 ms after
loading under changes in the focal length of the variable focus lens
in the range from 60 to 91 mm
with $\delta = -1.256$ THz for the detuning of the trap laser 
and 
15.6, 31.2, 46.8, and 62.4 mW of laser power, as shown in
Fig.~\ref{numexpcal} (a).
Each point represents a typical average over three samples, with 
the standard deviation as the error bar.
The observed temperatures in the NORT are typically below 10 $\mu$K.
The dependence of the number of trapped atoms on the focal length of the
variable focus lens shows a symmetricity centered around the
focal point in the trap laser beam.
When the laser power is high, 
there are two maximum peaks for the number of trapped atoms,
whereas only a single peak is observed for low power values of the laser.

We have numerically calculated the number of trapped atoms
using the cumulative distribution function of the
Maxwell-Boltzmann distribution for a temperature of 10 $\mu$K for
the atomic clouds, as expressed in Eq.~(\ref{eq7}).
Figure \ref{numexpcal} (b) shows the calculation results 
with one fitting parameter normalized to compensate for 
the average number of atoms in the two 
peaks in the experiment with 62.4 mW of the NORT intensity.
The trap efficiency is affected by gravity
owing to the shallow depth of the potential
in the branch side of the optical trap,
which reduces the number of trapped atoms in the branch
corresponding to a typical difference of 1 $\mu$K.
With respect to the potential of the NORT,
the radiative force generates the vector potential along the beam axis,
whereas the dipole force constructs the scalar potential gathering cold
atoms in the initial holding positions.
Owing to the characteristics of the vector potential in the NORT,
the potential depth in the radial direction remains unchanged, 
whereas that in the axial direction is modified.
The displacement of the atomic cloud is insensitive to 
the efficiency of atom trapping in the NORT.
The calculation results are in good agreement with the experiments, 
especially when the two maximum peaks can be explained as 
large trapping volumes that are dominant in the higher power regime.
In the case of low power,
the calculation results are comparable to those of the experiments for the
number of atoms in the trap,
although there are some differences for the dependence on focal length.
The potential depth of the trap in the low power regime
is sensitive to the beam profile of the trap, such as 
the spatial mode and tilted angle from the horizontal plane of the beam, 
which reduces the total trap depth generated by the dipole force.
The agreements between the calculations and experiments
indicate that 
the dipole force of the NORT determines 
the efficiency of initial atom trapping, whereas the radiative force has no such effect.
\begin{figure}[thbp]
\begin{center}
\includegraphics[width=7.0cm]{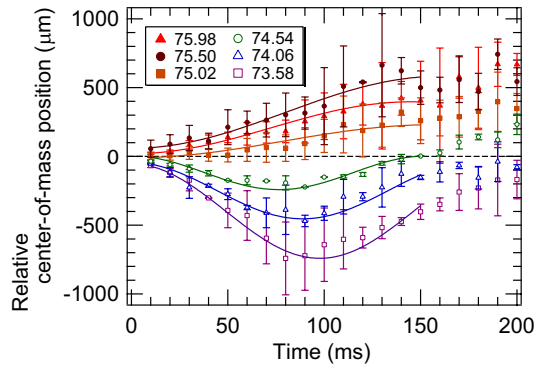}
\end{center}
\caption{(color online).
Trap time dependence of the center-of-mass positions of the trapped
atoms in the NORT.
The focal lengths of the variable focus lens are
75.98 mm (closed triangles),
75.50 mm (closed circles),
75.02 mm (closed squares),
74.54 mm (open circles),
74.06 mm (open triangles),
and 73.58 mm (open squares),
The lines are the fitting curves of the oscillation function obtained from Eq.~(\ref{eq8}).
} 
\label{comall}
\end{figure}

We have measured the time evolution of the trapped atomic cloud in the NORT. 
The trapped atoms start to move in the trap along the general axial direction.
The hybrid trap of the NORT that consists of both the radiative and dipole forces
produces an additional radiative force on the atoms as compared to the FORT beam.
Figure \ref{comall} illustrates the time evolution of the center-of-mass
positions of atomic clouds in the NORT for different focal lengths, namely 
75.98 mm, 75.50 mm, 75.02 mm, 74.54 mm, 74.06 mm, and 73.58 mm,
of the variable focus lens with a detuning of $-1.256$ THz and
trap laser power of 31.2 mW.
We define the evolution time of 0 ms to be just after the loading of the NORT
with the gray molasses.
The initial positions of the atoms are located at the center of the molasses.
The zero point of the relative center-of-mass position is defined as
the initial position for the focal length of 75.02 mm at the evolution time $t =$ 0 ms.
The relative center-of-mass positions of the atomic clouds
have displacements with oscillation-like behaviors.
When the focal lengths exceed 75 mm,
the atomic cloud  moves to the positive side with the beam propagation. 
However, when the length is less than 75 mm, the atoms  move to negative side.
We confirm that
there are limitations to the amplitude for the negative region
regardless of the fact that there are no limitations in the positive region.
Owing to the vector potential of the NORT, 
the radiative force deforms the trap potential along the beam axis.
In the region of the Rayleigh length,
the behaviors of atoms initially located at off-focus points can be regarded as a cosine function.
Therefore,
in Fig.~\ref{comall},
we fit the experimental data with the function 
\begin{equation}
z(t) = A {\rm cos}(\omega t + \theta) + z_1,
\label{eq8}
\end{equation}
where $\omega$ is the angular frequency of the oscillation, 
$z_1$ is the offset value of the oscillation,
and $A$ is the oscillation amplitude of the atomic cloud in the NORT.
We consider that 
the trapped atoms at the initial positions 
with approximately zero velocity
start oscillating in the NORT harmonically with the oscillator potential for $\theta = \pi$. 

We have studied 
the dependence of 
the amplitudes $A$ from Fig.~\ref{comall} on the focal length.
By linearly fitting the amplitudes $A$ depending on the focal length $f$,
we can find the equilibrium focal length at the zero-crossing point 
as $A = 0$ and then estimate $f =$ 75.08 mm for $\delta = -1.256$ THz.
In the case of $A = 0$, 
the atomic cloud tends to stay at the initial position 
because of the cancellation of the dipole and radiative forces of the NORT
in equilibrium condition.
The equilibrium focal length
is regarded as the most stable position to hold the atoms for a longer trap time.
We note that
we have fitted the amplitude in the range of 
more than a few millimeters away from the focus in the initial
position and confirmed larger amplitudes 
with non-harmonic oscillations rather than cosine oscillations.

\begin{figure}
\begin{center}
\includegraphics[width=6.5cm]{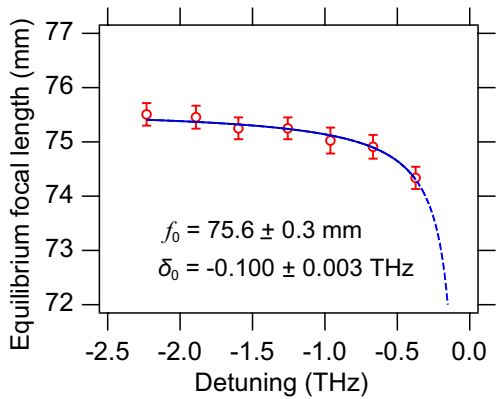}
\end{center}
\caption{(color online).
Detuning dependence of the equilibrium focal length in balance.
The solid curve is fitted using Eq.~(\ref{eq4}).
The dashed line is the extrapolated estimation using the fitting
parameters.
The fitting results show that the equilibrium focal length $f_0 = 75.6 \pm 0.3$ mm
and the threshold of detuning $\delta_0 = -0.100 \pm 0.003$ THz for trapping capability.
}
\label{eqfoc}
\end{figure}
Figure \ref{eqfoc} shows the 
equilibrium  focal length for each frequency of detuning
derived from estimation of the non-oscillation focal length with $A = 0$ 
for each trap time dependence of the center-of-mass positions 
as those in Fig.~\ref{comall}.
For the purpose of compensation for the same trap depth,
the laser power values of the NORT are set to $7.8$, $17.5$, $25.3$, $32.5$,
$43.4$, $52.2$, and $65.1$ mW, corresponding to the frequency detunings
$\delta = -0.373$, $-0.668$, $-0.962$, $-1.256$, $-1.598$, $-1.891$,
and $-2.232$ THz, respectively. 
The equilibrium focal lengths near resonance under $|\delta| < 1.0$
THz suddenly decrease with decrease of the absolute detuning $|\delta|$,
whereas those in $|\delta| > 1.0$ THz slightly decrease.
We fit the detuning dependence of the equilibrium focal length using 
Eq.~(\ref{eq4})
with $z_{\rm eq} = f_0 - f$, where $f_0$ is the asymptote focal length.
The asymptote to the value of $f_0 = 75.6 \pm 0.3$ mm is close to 
the focus of the trap beam in the far-off resonant regime.
We have confirmed that there is no equilibrium condition 
in threshold conditions
close to the resonance of $\delta \sim -0.15$ THz in our experiments, 
and the results are comparable to the fitting values of 
the divergence condition in the detuning case of $\delta_0 = -0.100 \pm 0.003$ THz.
The threshold for the divergence is determined by 
the confocal parameter $z_0$ and the detuning $\delta$,
as expressed by Eq.~(\ref{eq4}),
and it is insensitive to the laser power.

\begin{figure}
\begin{center}
\includegraphics[width=7.0cm]{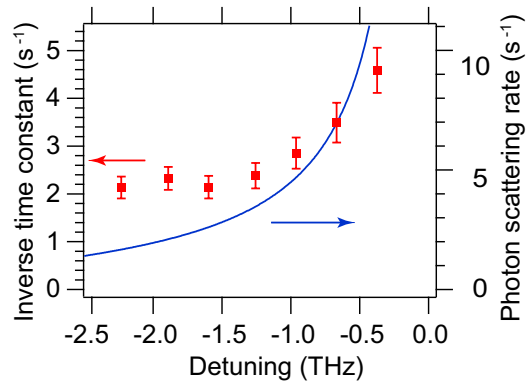}
\end{center}
\caption{(color online).
Detuning dependence of inversed lifetimes of trapped atoms in equilibrium conditions,
The solid line denotes the photon scattering rates of the NORT on the beam axis.
}
\label{lfscat}
\end{figure}

The radiative force in small detunings can generate
not only an additional force of the optical trap
but also the undesired heating effect caused by the photon scattering. 
Figure~\ref{lfscat} shows the detuning dependence of the inverse time
constants of the trapped atoms with the equilibrium focal conditions,
and that of the photon scattering rate of the NORT.
The experimental results are fitted by the single exponential function owing to
one body loss of the trapped atoms in our experimental condition \cite{Shibata}.
In the condition of $\delta <$ 1.0 THz,
the inverse time constant increases owing to 
the large effect of the radiative force, as shown in Fig.~\ref{lfscat}.
The detuning dependence of the inverse time constants is similar to 
that of the photon scattering rates.
Nevertheless, the quantitative differences
in the inverse of time constant in $\delta <$ 1.0 THz
are only a few factors of magnitude larger than those in $\delta <$ 1.0 THz.
The results indicate that the
NORT in the hybrid trap condition
near detuning from sub- to several THz
can be applied to ultracold experiments.

The ratio of 
the equilibrium displacement to the confocal parameter,
defined as $R = z_{\rm eq}/z_0 $,
informs us of the experimental conditions for realization of 
the hybrid trap system.
In our NORT system, 
we can estimate the experimental value of 
$R = 0.030$ with $z_{\rm eq} = 94$ $\mu$m at $\delta = -2.23$ THz,
and 
$R = 0.40$ with $z_{\rm eq} = 1.26$ mm at $\delta = -0.373$ THz.
The results are comparable to the calculation results:
$R = 0.068$ with $z_{\rm eq} = 215$ $\mu$m at $\delta = -2.23$ THz
and
$R = 0.51$ with $z_{\rm eq} = 1.61$ mm at $\delta = -0.373$ THz,
respectively.
The values of $R$ at the NORT are one or two orders of magnitude larger
than those in the FORT
for the same optical conditions;
{\it e.g.}, in the system using $\lambda = 1064$ nm, 
$R = 0.0015$ generates $z_{\rm eq} = 6.33$ $\mu$m 
at $\delta = -102$ THz in our trap system.
It is roughly estimated that 
the threshold for the realization of the hybrid trap regime 
is $R > 0.01$ 
for the condition where the range of the Rayleigh length
in a single trap is a few millimeters.
The NORT system has large advantages in comparison with the FORT system
because in large detunings
the value $R$ is approximately proportional to $1/|\delta|$ and
is independent of the laser power.
The sensitivity to the confocal parameter indicates that 
the radiative force can help the typical off-resonant trap with
additional degrees of freedom as the spatial mode,
such as the Laguerre-Gaussian mode \cite{Klimov,Shumyatsky}.
In addition,
the trap system used in the typical FORT with large $R$
can be realized with a large value of the confocal parameter;
{\it e.g.,}
$R > 0.1$ generates $z_{\rm eq} = 40.1$ mm with $z_0 = 339$ mm
in $\delta = -102$ THz with $f = 500$ mm and 1 mm of the beam radius.
Regardless of the large $|\delta|$ traps such as the FORT and the general optical tweezer,
the hybrid trap can be applied for precise optical control of not only atoms but also 
nanoscopic and mesoscopic media owing to the effect of the remaining
radiative force on the optical trap potential.

\section{Conclusion}

We have achieved a new method of hybrid optical trapping, which
consists of both the dipole and radiative forces, using the near
optical resonant regime called NORT.
We have measured the number of trapped atoms depending on the focal
length of the trap
and oscillation behaviors in the NORT depending on the
detuning of the trap laser with atomic resonance.
We have calculated the number of atoms using Gaussian beams with the
cumulative function of the Maxwell-Boltzmann distribution at finite temperature.
The calculated results are in good agreement with the experiments
and express the larger number of atoms located at off-focus positions.
We have estimated the equilibrium position using the hybrid trap
and confirmed its power-independent position by
comparisons between the experiments and calculations.
This hybrid trap has a threshold detuning of 
$|\delta| > k_z^2{w_{0}}^2 \Gamma/(8\pi^2)$ 
and is modified to the trap potential 
with displacements along the beam propagation axis,
which is insensitive to the trap laser beam power.
The novel optical tweezer system including the radiative force 
can be applied to precisely control nano- and micro-particles
and mesoscopic media
in use of multiple traps and higher spatial mode of the trap beams.

\begin{acknowledgements}

We would like to thank
K. Shibata, S. Takashima, and Y. Iizuka for their assistance in
constructing the early experimental setup.
This work was supported by 
the Matsuo Foundation,
the Research Foundation for Opto-Science and Technology,
Grants-in-Aid for Scientific Research (No. 15K05234)
from the Ministry of Education, Culture, Sports, Science, and Technology
of Japan, 
and the Chuo University Joint Research Grant.

\end{acknowledgements}


\begin{thebibliography}{99}
\bibitem{Grier}
D. G. Grier, Nature \textbf{424}, 810 (2003).
\bibitem{ChuPRL}
S. Chu, J. E. Bjorkholm, A. Ashkin, and A. Cable,
Phys. Rev. Lett. \textbf{57}, 314 (1986).
\bibitem{Miller}
J. D. Miller, R. A. Cline, and D. J. Heinzen,
Phys. Rev. A \textbf{47}, R4567, (1993).
\bibitem{Gustavson}
T. L. Gustavson, A. P. Chikkatur, A. E. Leanhardt, A. G\"olitz, S. Gupta,
D. E. Prichard, and W. Ketterle,
Phys. Rev. Lett. \textbf{88}, 020401 (2001).
\bibitem{AshkinOL}
A. Ashkin, J. M. Dziedzic, J. E. Bjorkholm, and S. Chu,
Opt. Lett. \textbf{11}, 288 (1986).
\bibitem{Svobada}
K. Svobada and S. M. Block,
Opt. Lett. \textbf{19}, 930 (1994).
\bibitem{AshkinSci}
A. Ashkin and J. M. Dziedzic,
Science \textbf{235}, 4795 (1987).
\bibitem{AshkinPNAS}
A. Ashkin,
Proc. Natl. Acad. Sci.\textbf{94}, 4853 (1997).
\bibitem{MahamdehOL}
M. Mahamdeh and E. Sch\"affer,
Opt. Express \textbf{17}, 17190 (2009).
\bibitem{Metcalf}
H. Metcalf and P. van der Straten,
{\it Laser Cooling and Trapping} 
(Springer Verlag, New York, USA, 1999).
\bibitem{GrimmAMOP}
R. Grimm, M. Weidemuller, and Y. B. Ovchinnikov,
Adv. At. Mol. Opt. Phys. \textbf{42}, 95 (2000).
\bibitem{TojoPRA}
S. Tojo, T. Hayashi, T. Tanabe, T. Hirano, Y. Kawaguchi, H. Saito, and
M. Ueda,
Phys. Rev. A \textbf{80}, 042704 (2009).
\bibitem{StamperKurnRMP}
D. M. Stamper-Kurn and M. Ueda,
Rev. Mod. Phys. \textbf{85}, 1191 (2013).
\bibitem{PRLGaunt}
A. L. Gaunt, T. F. Schmidutz, I. Gotlibovych, R. P. Smith, and
Z. Hadzibabic,
Phys. Rev. Lett \textbf{110}, 200406 (2013).
\bibitem{NJPLeonard}
J. Leo\'nard, M. Lee, A. Morales, T. M. Karg, T. Esslinger, and T. Donner,
New J. Phys. \textbf{16}, 093028 (2014).
\bibitem{Szczepkowicz}
A. Szczepkowicz, L. Krzemien\', A. Wojciechowski, K. Brzozowski,
M. Kru\"ger, M. Zawada, M. Witkowski, J. Zachorowski, and W. Gawlik,
Phys. Rev. A \textbf{79}, 013408 (2009).
\bibitem{Bienaime}
T. Bienaime\', S. Bux, E. Lucioni, Ph. W. Courteille, N. Piovella, 
and R. Kaise,
Phys. Rev. Lett. \textbf{104}, 183602 (2010).
\bibitem{Mandel}
L. Mandel and E. Wolf,
{\it Optical coherence and Quantum optics} 
(Cambrdige University Press, New York, USA, 1995).
\bibitem{Lett}
P. D. Lett, W. D. Phillips, S. L. Rolston, C. E. Tanner, R. N. Watts,
and C. I. Westbrook, 
J. Opt. Soc. Am. B \textbf{6}, 2084 (1989). 
\bibitem{Dumke}
R. Dumke, M. Johanning, E. Gomez, J. Weinstein, K. Jones, and P. Lett,
New J. Phys. \textbf{8}, 64 (2006).
\bibitem{Shibata}
K. Shibata, S. Yonekawa, and S. Tojo,
Phys. Rev. A \textbf{96}, 013402 (2017).
\bibitem{Hemmierich}
A. Hemmerich, M. Weidemu\"ler, T. Esslinger, C. Zimmermann, and 
T. Ha\"nsch,
Phys. Rev. Lett. \textbf{75}, 35 (1995).
\bibitem{Boiron}
D. Boiron, A. Michaud, P. Lemonde, Y. Castin, C. Salomon,
S. Weyers. K. Szymaniec, L. Gognet, and A. Clairon,
Phys. Rev. A \textbf{53}, R3734 (1996).
\bibitem{Klimov}
V. V. Klimov, D. Bloch, M. Ducloy, and J. R. Rios Leite,
Phys. Rev. A \textbf{85}, 053834 (2012).
\bibitem{Shumyatsky}
P. Shumyatsky, G. Milione, and R. R. Alfano,
Opt. Commun. \textbf{321}, 116 (2014).

\end{thebibliography}
\end{document}